\documentclass[11pt,twoside]{article}

\usepackage{asp2006}
\usepackage{epsf}
\usepackage{psfig}
\usepackage{lscape}

\begin{document}
\title{A hole in the sky -- The dependence of the galaxy luminosity function on the environment}   
\author{S. Phleps, C. Wolf, J.A. Peacock, K. Meisenheimer, and
  E. van Kampen}   
\affil{}    

\begin{abstract} 
We have developed a method to calculate overdensities in multicolour
surveys, and compare the local density contrast
measured in galaxy samples with different redshift error
distributions. We calculated overdensities for three COMBO-17 fields,
and identified a region in the CDFS, where the density is lower by a
factor of 2 compared to the other two fields. This is mainly due to a
deficiency of faint red galaxies. This result is in
agreement with local observations in the 2dF.
\end{abstract}
\begin{figure}[h]
\centerline{\psfig{figure=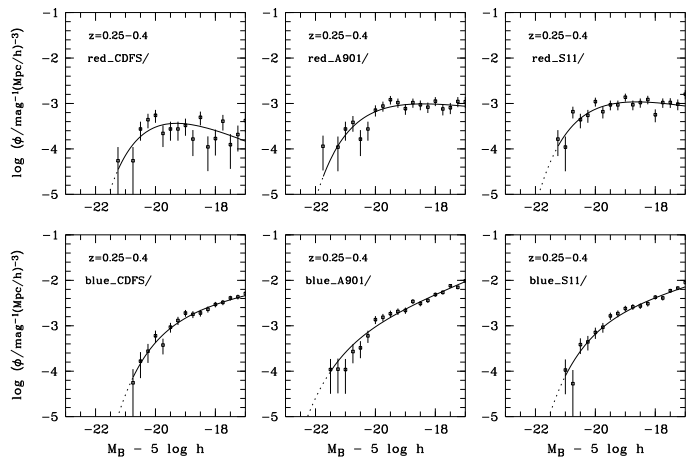,angle=270,clip=t,width=10.cm}}
\caption[ ]{The luminosity function of the red sequence (upper panel)
and blue cloud (lower panel) galaxies in
the redshift range $0.25 \leq z \leq 0.4$,
 for the three COMBO-17 fields.
The STY fit is overplotted in each panel.\label{lumifunc}}
\end{figure}

It is a well established fact that galaxy properties such as
morphology and colour depend on environment -- red, early-type
galaxies are usually found in denser regions of the universe, whereas
the field is dominated by blue, late-type galaxies. The dependence of
the galaxies' appearance on the local density is a clue to their
formation and evolution, so it is important to investigate it in some
detail. A powerful means to do that is the type-dependent luminosity
function (LF in the following), calculated in different overdensity regimes.
\begin{figure}[b]
\centerline{\psfig{figure=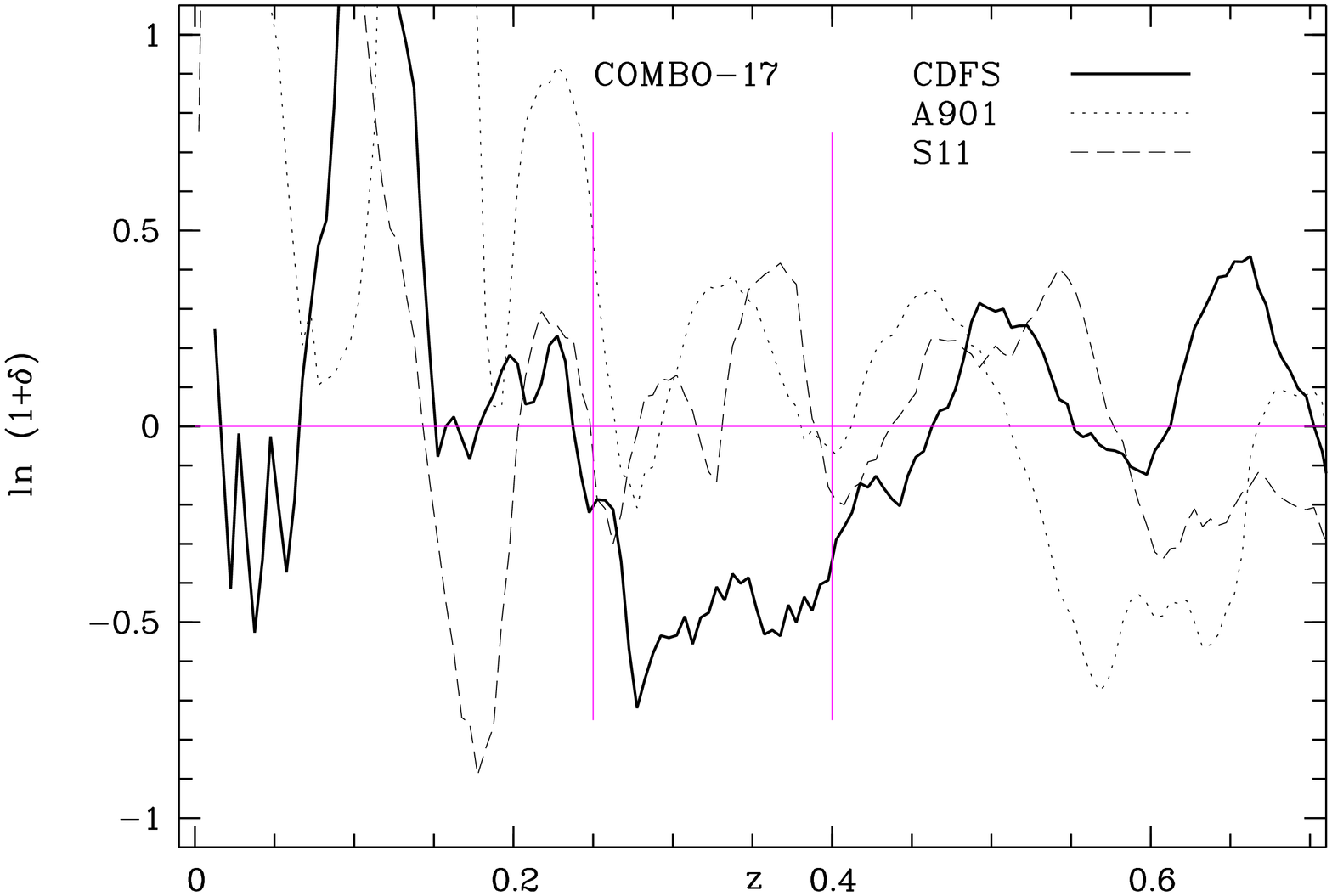,angle=270,clip=t,width=5.cm}\psfig{figure=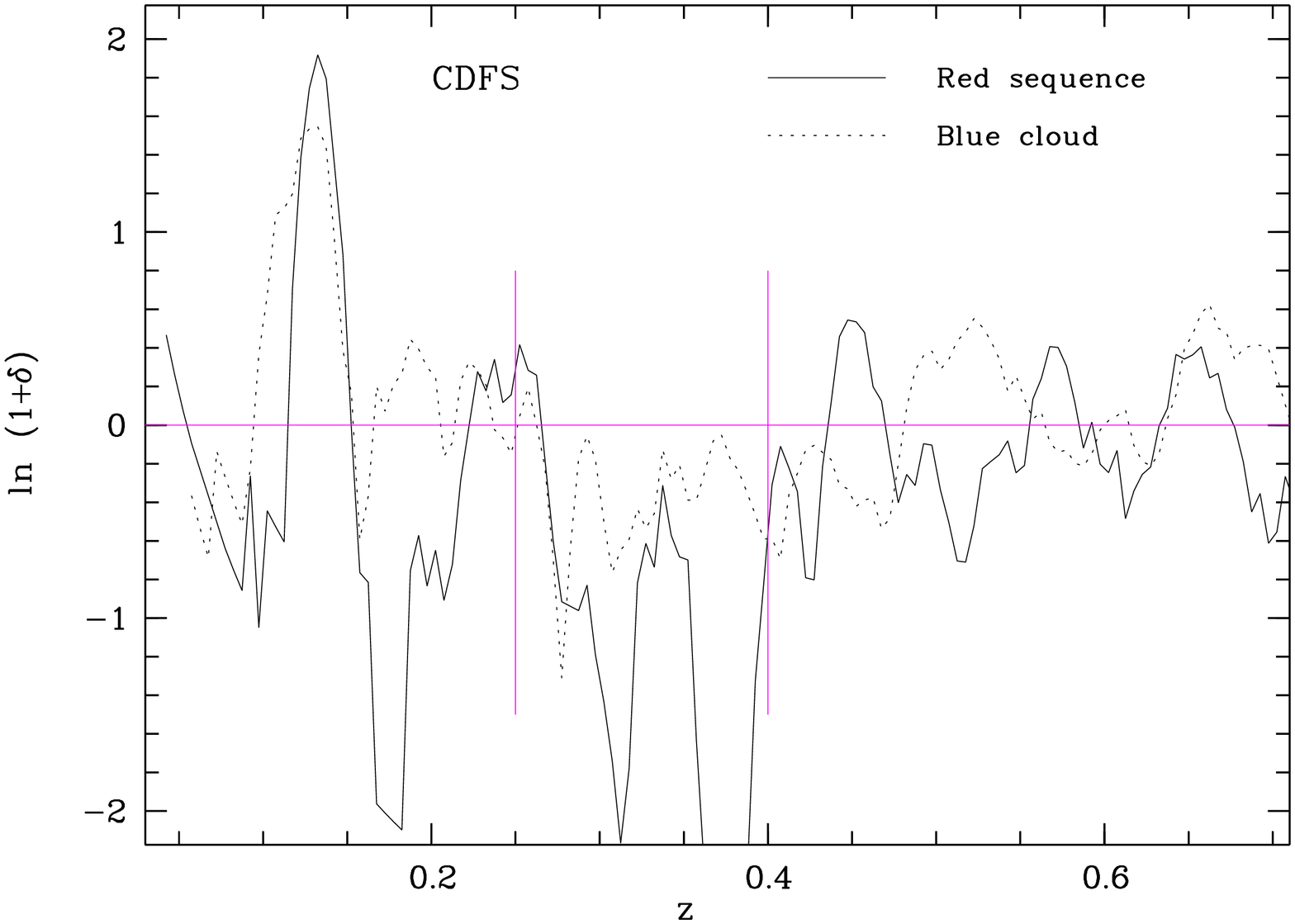,angle=270,clip=t,width=5.cm}}
\caption[ ]{{\bf Left:} Overdensities in three COMBO-17 fields. {\bf
    Right:} For red and blue galaxies in the CDFS. The vertical lines
    indicate the range in which the LF is calculated.
\label{densities}}
\end{figure}
With spectroscopic data it is possible to measure the local galaxy
density contrast $\delta_g=(\rho/\bar{\rho_0})-1$ in small spheres of
e.g. $8 h^{-1}$\,Mpc. However, for an analysis at
higher redshifts, deep multicolor
surveys have to be used, and those have too large redshift
inaccuracies to facilitate a determination of $\delta_g$ in small
spheres. Instead, $\delta_g$ can be calculated  in small redshift bins
(with binsize $\Delta z=0.02$ and in steps of
$\delta_z=0.005$). We tested the influence of redshift errors on the
determination of $\delta_g$ using a COMBO-17 mock survey
\citep{vanKampen05}, and found that the structures are smoothed
according to the size of the errors. Red galaxies, which are
statistically brighter, have smaller errors than the (statistically)
fainter blue galaxies, so in order to directly compare
the two samples, we have to calculate a blurring function which renders
their redshifts as inaccurate as the blue ones.  The blurring function
can be found via the convolution  
theorem and then applied to the data in a Monte-Carlo fashion. 
We used COMBO-17 data (see \citealp{Wolf03})
and calculated the overdensities for all galaxies with $R\leq 23.65$
and rest-frame $B$-band magnitudes $M_B\leq-18.0$, see left panel of
Fig. \ref{densities}. In the redshift range $0.25 \leq z \leq 0.4$
the CDFS (Chandra Deep Field South) is underdense with respect to the other two fields, where the
underdensity fluctuates about the mean. We split the sample into red
sequence and blue cloud galaxies (see \citealp{Bell04}) and repeated the
calculation after blurring the red galaxies' redshifts. In the
calculation of the mean density $\bar\rho$ the increasing number
density of blue galaxies  with redshift has to be taken into account, which we
estimate from a fit to the measured mean in the three fields. The
right panel of Fig. \ref{densities} shows the resulting underdensities
in the CDFS. The underdensity at $0.25 \leq z \leq 0.4$ is clearly
deeper in the red sample.
We calculated the type-dependent LF in this redshift
range, for each field separately, see Fig. \ref{lumifunc}.
While the LF of the blue galaxies remains unaffected by environment,
the LF of the red galaxies does depend on the local density: In the
CDFS the faint-end slope is clearly more positive than in the other
fields. Hence, the 'hole in the sky' is mainly due to a deficiency of
faint red galaxies.
This result is in general agreement with the result of
\citet{Croton05}, who found the LF of blue 2dF galaxies to be essentially constant
with varying overdensity, and the faint-end slope of the LF of the red
galaxies to be steeper in higher density regions.





\end{document}